\documentclass[prx,twocolumn,superscriptaddress,citeautoscript,amsart,longbibliography,nofootinbib]{revtex4-2}

\usepackage{graphicx}
\usepackage{multirow}
\usepackage{color}
\usepackage{bm}
\usepackage{times}
\usepackage{amsmath,bm,amsfonts}
\usepackage{dcolumn}
\usepackage{graphicx}
\usepackage{latexsym}
\usepackage{hhline}
\usepackage{braket}
\usepackage{bbold}
\usepackage{multirow}

\def\ket#1{|#1\rangle }

\def\bb{\mathbb}
\def\Pf{{\rm Pf}}
\def\d{\partial}

\begin{document}
\title{
Many-body selection rule for quasiparticle pair creations in centrosymmetric superconductors
}
\author{Junyeong \surname{Ahn}}
\email{junyeongahn@fas.harvard.edu}
\affiliation{Department of Physics, Harvard University, Cambridge, MA 02138, USA}

\author{Naoto \surname{Nagaosa}}
\email{nagaosa@riken.jp}
\affiliation{RIKEN Center for Emergent Matter Science (CEMS), Wako, Saitama 351-0198, Japan}
\affiliation{Department of Applied Physics, The University of Tokyo, Bunkyo, Tokyo 113-8656, Japan}

\date{\today}

\begin{abstract}
When metal becomes superconducting, new optical excitation channels are created by particle-hole mixing.
These excitation channels contribute negligibly to optical responses in most superconductors, but they can be relevant in ultra-strong-coupling superconductors that are close to the Bose-Einstein condensate regime.
Recently, selection rules for these excitations have been formulated based on single-particle anti-unitary symmetries in the mean-field theory.
While being potentially useful for studying optical properties of ultra-strong-coupling superconductors, they had fundamental limitations because significant quantum fluctuations invalidate mean-field approaches.
Here, we use many-body states to formulate an optical selection rule that does not rely on the mean-field approximation.
In this approach, the physical meaning of the previous selection rules becomes clearer as they are simply recast as the selection rule for many-body inversion eigenstates, not involving anti-unitary symmetries.
This selection rule applies not only to the Bogoliubov quasiparticles of Fermi liquids but also to non-Fermi-liquid quasiparticles and electrically charged bosonic excitations.
We also study the Bogoliubov Fermi surfaces, whose topological stability is closely related to the selection rule.
We provide a many-body formulation of their topological charges and show that the low-energy optical conductivity of the Bogoliubov Fermi surfaces depends crucially on their secondary topological charge.
Finally, we discuss the implications of our results to the stability of the superconducting state.
\end{abstract}

\maketitle

\section{Introduction}

Conventional Bardeen-Cooper-Schrieffer (BCS) superconductors respond to static electromagnetic fields as if almost all conduction electron pairs contribute to the supercurrent~\cite{tinkham2004introduction}.
This property follows from the optical property of the BCS model:
Cooper pairs are not broken when a spatially uniform electric field is turning on because their optical spectral weight is zero.
The absence of optical excitations in the BCS model has been attributed to the crystal momentum conservation imposed by translation symmetries~\cite{mattis1958theory,mahan2013many}.

It has been recently found that an optical selection rule imposed by additional inversion symmetry is necessary for the absence of pair-breaking low-energy optical excitations~\cite{xu2019nonlinear,ahn2021theory}.
The selection rule was formulated within the mean-field Bogoliubov-de Gennes (BdG) formalism, where the relevant symmetry operation was the single-particle anti-unitary operation defined by the combination of inversion and particle-hole conjugation~\cite{ahn2021theory}.
The establishment of this selection rule provided routes to pair-breaking optical responses in single-crystal superconductors.
Xu, Morimoto, and Moore showed that Cooper pairs can be optically broken in superconductors without inversion symmetry~\cite{xu2019nonlinear}.
Ahn and Nagaosa showed further that multiband effects can allow for the optical pair breaking even in the presence of inversion symmetry~\cite{ahn2021theory}.

However, the applicability of this inversion selection rule in real superconductors is in question.
The optical transition probability of a Cooper-pair breaking accompanies the factor of $(k_F\xi)^{-2}$, where $k_F$ is the Fermi wave vector and $\xi$ is the superconducting coherence length, which is typically very small.
This factor can be non-negligible only when the Thomas-Fermi screening length $\sim k_F^{-1}$ is comparable to the coherence length, in which case quantum fluctuations give a significant correction to the mean-field results.
Since the selection rule in Refs.~\cite{xu2019nonlinear,ahn2021theory} is formulated within the mean-field theory, it is not clear whether it applies to ultra-strong-coupling superconductors close to the Bose-Einstein condensate (BEC) regime $(k_F\xi<1)$.
Addressing this issue is relevant for the optical study of the BCS-BEC crossover~\cite{chen2005bcs,randeria2014crossover} recently observed in superconductors, including iron-based superconductors~\cite{lubashevsky2012shallow,hashimoto2020bose,shibauchi2020exotic}, twisted bilayer graphene~\cite{cao2018unconventional}, twisted trilayer graphene~\cite{park2021tunable,hao2021electric}, and Li$_x$ZrNCl~\cite{nakagawa2021gate}.

Another issue in the mean-field approach is the U(1) gauge symmetry breaking by Cooper pairing, which leads to the non-conservation of electric charges in the superconducting state.
This may leave some anomalies in physical responses such as those related to violation of the sum rule~\cite{lin2018towards}.
A charge-conserving approach is desirable to establish an unambiguous result, especially to relate the optical conductivity to superfluid stiffness through the Ferrel-Glover-Tinkham (FGT) sum rule~\cite{ferrell1958conductivity,tinkham1959determination}.

\begin{figure}[t!]
\includegraphics[width=8.5cm]{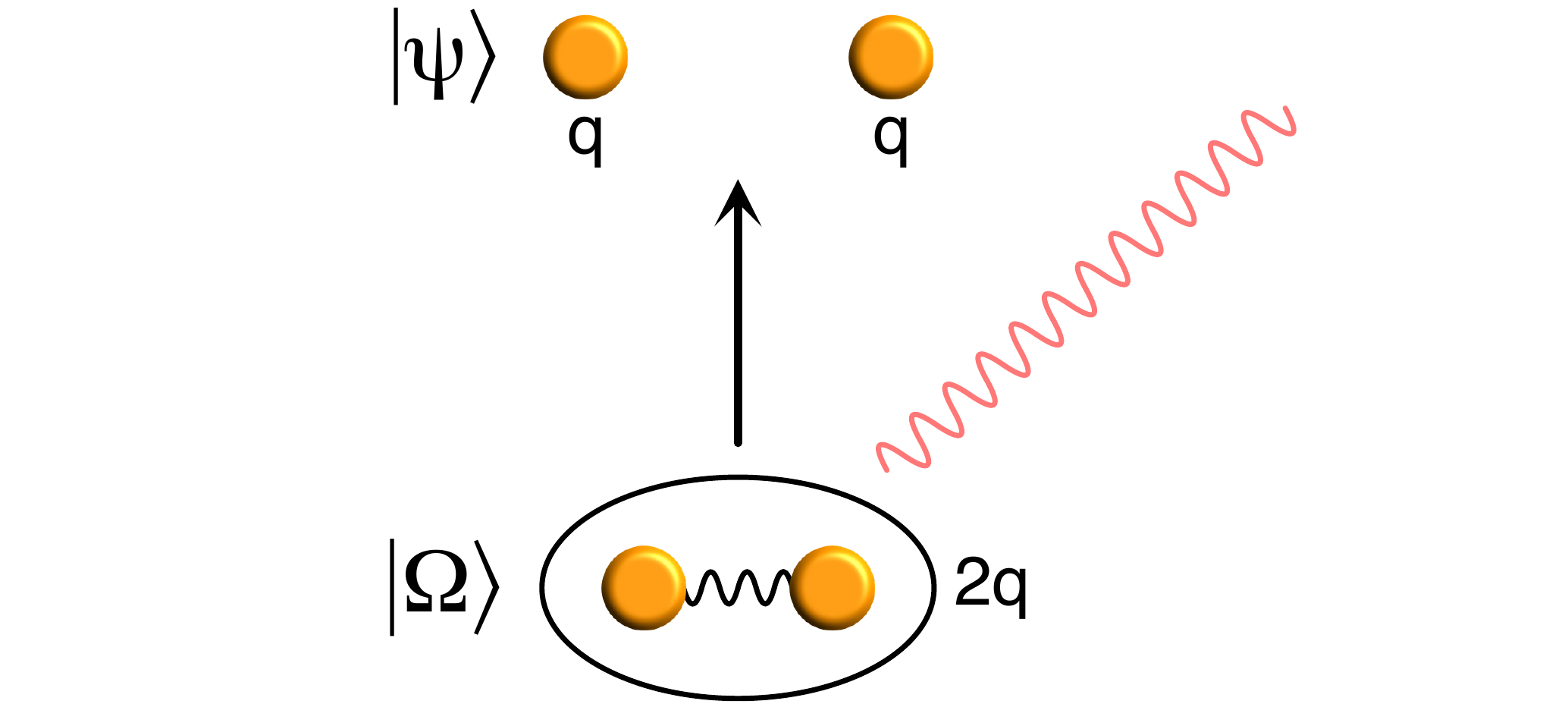}
\caption{
Optical breaking of a Cooper pair.
$q$ is the electric charge of a quasiparticle.
We consider the selection rule imposed on the optical process where a Cooper pair in a many-body state $\ket{\Omega}$ is excited to $\ket{\psi}$ with a broken pair of quasiparticles.
}
\label{fig:pair-breaking}
\end{figure}

In this paper, we promote the results of Refs.~\cite{xu2019nonlinear,ahn2021theory} to a form that applies beyond the mean-field theory.
We reformulate the inversion selection rule using many-body states in the charge-conserving theory in Sec.~\ref{sec:selection-charge-conserving}.
Our selection rule has the advantage of being physically clearer because it does not involve an anti-unitary operator.
This is possible because the particle-hole symmetry of the BdG Hamiltonian appears due to the redundancy in the formalism and is not a symmetry of the many-body Hamiltonian.
Since there is no particle-hole symmetry in the many-body approach, we need to consider inversion symmetry only, which imposes ordinary selection rules as a unitary symmetry.
Namely, we use that, for eigenstates $\ket{\Omega}$ and $\ket{\psi}$ of the many-body inversion operator $\hat{I}$ with eigenvalues $\lambda_{\psi}$ and $\lambda_{\Omega}$,
\begin{align}
\label{eq:selection-general}
\braket{\psi|\hat{\bf J}|{\Omega}}
=0
\text{ when }
\lambda_{\psi}\ne-\lambda_{\Omega},
\end{align}
where ${\bf \hat{J}}$ is the many-body current operator.
Hereafter we put the hat symbol on operators acting on the many-body Hilbert space.
We show below that the selection rules in Refs.~\cite{xu2019nonlinear,ahn2021theory} are reproduced when $\ket{{\Omega}}$ is the ground state and $\ket{\psi}$ is the excited state with inversion-related electron pairs.
This approach applies to non-Fermi liquid and bosonic quasiparticles also.
We connect our result to the previous mean-field single-particle approach in Sec.~\ref{sec:selection-MFT} and give some examples of the consequeces of the selection rule in Sec.~\ref{sec:examples}.

Our selection rule is intimately related to the topological stability of the Bogoliubov Fermi surfaces~\cite{ahn2021theory}, which also require the symmetry under the combination of particle-hole conjugation and inversion in the single-particle approach.
This suggests an interesting optical properties of Bogoliubov Fermi surfaces.
In Sec.~\ref{sec:BFS}, we reveal low-energy divergent optical responses of Bogoliubov Fermi surfaces carrying two topological charges.
Also, we show how to express the topological charges of Bogoliubov Fermi surfaces without involving particle-hole symmetry explicitly, by using a momentum-dependent many-body ground state.
We conclude in Sec.~\ref{sec:discussion} by discussing the stability of unconventional superconductors in light of our selection rule.

Throughout the paper, we assume that the external gauge field is uniform in space.
This neglects the effect of photon momentum ${\bf q}$ that comes as a small correction of the factor $q^2/k_F^2= (v_F/c)^2(\hbar qc/E_F)^2$, where $k_F$, $v_F$, and $E_F=\hbar v_Fk_F$ are Fermi wave vector, velocity, and energy, $c$ is the velocity of light, and $\hbar qc$ is the photon energy.

\section{Selection rule in charge-conserving theory}
\label{sec:selection-charge-conserving}

We first introduce the many-body inversion selection rule without mean-field approximation.
Let us recall that the optical conductivity tensor is expressed by the Kubo formula as
\begin{align}
\sigma^{ab}(\omega)
&=\frac{\hbar}{V}\sum_{m,n}\frac{\braket{n|\hat{J}^a|m}\braket{m|\hat{J}^b|n}}{E_m-E_n}\frac{-i(\rho_n-\rho_m)}{E_m-E_n-\hbar\omega}\notag\\
&\quad+\frac{ie^2}{\omega V}\left\langle\frac{\d^2\hat{H}}{\d A^a\d A^b}\right\rangle,
\end{align}
where $V$ is the volume of the system, $\hat{H}\ket{n}=E_n\ket{n}$ for the Hamiltonian $\hat{H}$, $\rho_n=\braket{n|\hat{\rho}|n}$ is the density matrix element, and $\hat{\bf J}=\partial\hat{H}/\partial {\bf A}$ is the current operator, and ${\bf A}$ is the external gauge field, and $\braket{\hat{O}}=\sum_n\braket{n|\hat{O}|n}\rho_n$ in the last line.
We are interested in the resonant transitions between two different energy levels, so the diamagnetic term in the second line is irrelevant.
We study the constraint of inversion symmetry on the transition amplitude $\braket{m|\hat{J}^a|n}$ for specific channels that we now explain.
In this section, we do not assume lattice translation symmetries.

We are interested in the excitation of a pair of quasiparticles created by $\hat{\gamma}^{\dagger}$ with charge $q$ in the superconducting state, where the Cooper pair carries charge $2q$.
In the simplest case, $\hat{\gamma}^{\dagger}$ is the electron creation operator and the Cooper pair is formed by two electrons, but we also consider non-fermionic quasiparticle excitations.
The excited state of our interest has the form
\begin{align}
\label{eq:two-particle}
\ket{\psi_{\alpha}}=\hat{\gamma}^{\dagger}_{\alpha}(\hat{I}\hat{\gamma}^{\dagger}_{\alpha }\hat{I}^{-1})\hat{\cal C}\ket{{\Omega}},
\end{align}
where $\ket{\Omega}$ is an inversion eigenstate, $\alpha$ is the collective index containing all orbital, spin, and real-space position or crystal momentum, $\hat{I}:(t,{\bf x})\rightarrow (t,-{\bf x})$ is an inversion operator, and the Cooper pair annihilation operator $\hat{\cal C}$ is needed to conserve the charge in the optical excitation.
We suppose that $[\hat{H},\hat{\gamma}^{\dagger}_{\alpha}]=E_{\gamma,\alpha}\hat{\gamma}^{\dagger}_{\alpha}$, $[\hat{H},\hat{\cal C}^{\dagger}]=E_{\cal C}\hat{\cal C}^{\dagger}$, and $[\hat{H},\hat{I}]=0$.
Then, if $\ket{\Omega}$ is an energy eigenstate, then $\ket{\psi_{\alpha}}$ is also an energy eigenstate.
Although the excitation of inversion-asymmetric pairs is also possible in centrosymmetric systems, we are interested in the excitation of inversion-related pairs [Eq.~\eqref{eq:two-particle}] because this channel can be the lowest pair excitation if it is not forbidden by the selection rule~\cite{ahn2021theory}.
We consider arbitrary spatial dimensions.
It is convenient to define $\hat{I}$ such that its square may contain an overall U(1) phase rotation, i.e., $\hat{I}^2\hat{\gamma}^{\dagger}_{\alpha}\hat{I}^{-2}=\epsilon_I\hat{\gamma}^{\dagger}_{\alpha}$, where $\epsilon_I=\pm 1$, although this phase can be removed by re-defining $\hat{I}$ to contain a phase rotation.
For example, the nontrivial factor $\epsilon_I=-1$ can arise due to spin $2\pi$ rotation of fermions when $\hat{I}$ represent the spatial inversion times a spin $\pi$ rotation in three dimensions or the twofold rotation around the normal axis in two dimensions.

Equation~\eqref{eq:selection-general} means that an inversion eigenstate $\ket{\Omega}$ with the eigenvalue $\lambda_{\Omega}$ can be excited only to those states with inversion eigenvalue $-\lambda_{\Omega}$ because the current operator has odd parity under inversion.
Let us evaluate the inversion eigenvalue of $\ket{\psi_{\alpha}}$.
We suppose that $\hat{\gamma}^{\dagger}$ satisfies the commutation relation
\begin{align}
\label{eq:condition1}
\hat{\gamma}^{\dagger}_{\alpha}\hat{\gamma}^{\dagger}_{\beta}
&=\eta_{\gamma}\hat{\gamma}^{\dagger}_{\beta}\hat{\gamma}^{\dagger}_{\alpha},
\end{align}
where $\eta_{\gamma}=\pm 1$, and transforms under $\hat{I}$ as 
\begin{align}
\label{eq:condition2}
\hat{I}\hat{\gamma}^{\dagger}_{\alpha}\hat{I}^{-1}
=\hat{\gamma}^{\dagger}_{\beta}(V_I)_{\beta\alpha},
\quad (V_I^2)_{\beta\alpha}=\epsilon_I\delta_{\beta\alpha}.
\end{align}
Note that $\hat{\gamma}^{\dagger}_{\alpha}$ may not be a fermionic or bosonic creation operator, because we do not require canonical commutation relations $\hat{\gamma}_{\alpha}\hat{\gamma}^{\dagger}_{\beta}+\eta_{\gamma}\hat{\gamma}^{\dagger}_{\beta}\hat{\gamma}_{\alpha}= \delta_{\alpha\beta}$.
It follows from Eqs.~\eqref{eq:condition1} and~\eqref{eq:condition2} that
$
\hat{I}[\hat{\gamma}^{\dagger}_{\alpha}(\hat{I}\hat{\gamma}^{\dagger}_{\alpha }\hat{I}^{-1})]\hat{I}^{-1}
=\eta_{\gamma}\epsilon_I\hat{\gamma}^{\dagger}_{\alpha}(\hat{I}\hat{\gamma}^{\dagger}_{\alpha }\hat{I}^{-1}).
$
The eigenvalues of the Cooper pair and the initial state $\ket{\Omega}$ are
\begin{align}
\hat{I}\hat{\cal C}^{\dagger}\hat{I}^{-1}
&=\lambda_{\cal C}\hat{\cal C}^{\dagger},\notag\\
\hat{I}\ket{\Omega}
&=\lambda_{\Omega}\ket{\Omega},
\end{align}
where $\lambda_{\cal C}=\pm 1$ and $\lambda_{\Omega}=\pm 1$.
The inversion eigenvalue of $\ket{\psi_{\alpha}}$ is then $\eta_{\gamma}\epsilon_I\lambda_{\cal C}\lambda_{\Omega}$, so we arrive at our main result.
\begin{align}
\label{eq:selection-rule}
\braket{\psi_{\alpha}|\hat{\bf J}|{\Omega}}
=0
\text{ when }
\lambda_{\cal C}
\ne -\eta_{\gamma}\epsilon_I.
\end{align}
Note that $\lambda_{\Omega}$ does not appear in the final expression.

This selection rule is intimately related to the energy spectrum of the two-particle state $\ket{\psi_{\alpha}}$.
Let us consider the case where $\ket{\Omega}$ is the superconducting ground state.
When $\ket{\psi_{\alpha}}$ and $\ket{\Omega}$ have the same inversion eigenvalue, there is band repulsion between them.
Therefore, the spectrum of $\ket{\psi_{\alpha}}$ is gapped in general.
More precisely, more than three parameters (contained in $\alpha$ in our notation) need to be adjusted to close the gap.
For example, in three-dimensional momentum space, the gap can close at isolated points, generically. 
The density of states of normal charge carriers vanishes at zero temperature.
On the other hand, when $\ket{\psi_{\alpha}}$ and $\ket{\Omega}$ have the opposite inversion eigenvalues, no band repulsion occurs between them.
It is possible to close the gap by tuning only one parameter.
Therefore, zero-energy states with a non-vanishing density of states can appear generically (although not necessarily~\footnote{It is possible to have a gapped spectrum while having a nonzero oscillator strength for pair productions, $\braket{\psi_{\alpha}|\hat{\bf J}|{\Omega}}\ne 0$, in the clean limit.
This is interesting because Bogoliubov quasiparticles behave like neutral quasiparticles due to perfect charge screening in superconductors, which makes one expect that Bogoliubov quasiparticles have zero oscillator strength~\cite{kivelson1990bogoliubov}.}).
In mean-field superconductors, this means the formation of the so-called Bogoliubov Fermi surface~\cite{agterberg2017bogoliubov} where $\gamma^{\dagger}_{\alpha}$ is the Bogoliubov quasiparticle creation operator.
As the appearance of such extensive gapless states is energetically unfavorable in most cases, it is typical to have $\hat{I}\ket{\psi_{\alpha}}=\lambda_{\Omega}\ket{\psi_{\alpha}}$ such that $\braket{\psi_{\alpha}|\hat{\bf J}|{\Omega}}=0$.

\section{Selection rule in mean-field theory}
\label{sec:selection-MFT}

As the selection rule is based on symmetry, one might expect that mean-field theory also gives the same result.
Here we re-derive the above selection rule within the mean-field theory of electronic quasiparticles and reproduce the results in Refs.~\cite{xu2019nonlinear,ahn2021theory} based on the single-particle approach.
Mean-field theory assumes the nontrivial expectation value of the superconducting order parameter
\begin{align}
b_{\alpha\beta}
&=\braket{{\Omega}^G_{\rm MF}|\hat{c}_{\alpha}\hat{c}_{\beta}|{\Omega}^G_{\rm MF}},
\end{align}
where $\hat{c}_{\alpha}$ is the electronic quasiparticle annihilation operator.
A nonzero value of the order parameter means that the mean-field ground state $\ket{\Omega^G_{\rm MF}}$ breaks the charge conservation.
Also, when the order parameter has odd parity, $\ket{\Omega^G_{\rm MF}}$ breaks the inversion symmetry of the normal state, while it preserves the symmetry under the combination of inversion and a phase transformation by $\pm \pi/2$.
Namely, when the order parameter satisfies
\begin{align}
(U_I)_{\alpha\gamma}b_{\gamma\delta}(U_I^T)_{\delta\beta}
=e^{2i\phi}b_{\alpha\beta},
\end{align}
where $T$ in the superscript is the matrix transpose, the combination
\begin{align}
\hat{I}_{\phi}=\hat{U}_{\phi}\hat{I}
\end{align}
remains a symmetry operator, where
\begin{align}
\hat{U}_{\phi}\hat{c}_{\alpha}\hat{U}_{\phi}^{-1}
&=e^{i\phi}\hat{c}_{\alpha},\notag\\
\hat{I}\hat{c}_{\alpha}\hat{I}^{-1}
&=\hat{c}_{\beta}(U_I)^*_{\beta\alpha},
\end{align}
and $e^{2i\phi}=\pm 1$.
In comparison, note that the superconducting ground state preserves inversion symmetry and particle number conservation in the charge-conserving formalism.

The two-particle-excited state we consider in the mean-field theory is thus
\begin{align}
\label{eq:MF-two-particle}
\ket{\psi_{\alpha}^{\rm MF}}=\hat{\gamma}^{\dagger}_{\alpha}(\hat{I}_\phi\hat{\gamma}^{\dagger}_{\alpha}\hat{I}_\phi^{-1})\ket{{\Omega}_{\rm MF}},
\end{align}
where $\ket{{\Omega}_{\rm MF}}$ is an eigenstate of $\hat{I}_\phi$, instead of $\hat{I}$, where $\hat{\gamma}^{\dagger}$ is the Bogoliubov quasiparticle creation operator.
We do not need to include the Cooper pair annihilation operator $\hat{\cal C}$ here because particle number conservation is broken here.
In fact, the selection rule does not depend on how many $\hat{\cal C}$ operators we include, since we formulate the selection rule in terms of $\hat{I}_{\phi}$, under which the Cooper pair is invariant, i.e., $\hat{I}_{\phi}\hat{\cal C}\hat{I}_{\phi}^{-1}=\hat{\cal C}$.
As we assume $\hat{I}_\phi$ symmetry, a Bogoliubov quasiparticle is transformed to another Bogoliubov quasiparticle under $\hat{I}_\phi$: $\hat{I}_{\phi}\hat{\gamma}^{\dagger}_{\alpha}\hat{I}_{\phi}^{-1}=\hat{\gamma}^{\dagger}_{\beta}(V_{I_{\phi}})_{\beta\alpha}$.
Using that $\hat{I}_{\phi}{\bf J}\hat{I}_{\phi}^{-1}=-{\bf J}$ and following the same steps in the previous section, we derive
\begin{align}
\label{eq:MF-selection-rule}
\braket{\psi_{\alpha}^{\rm MF}|\hat{\bf J}|{\Omega}_{\rm MF}}
=0
\text{ when }
e^{2i\phi}
\ne -\eta_{\gamma}\epsilon_I.
\end{align}
One can see that this is exactly the selection rule in Eq.~\eqref{eq:selection-rule} by noting that the inversion eigenvalue of the order parameter, $e^{2i\phi}$, is the same as that of the Cooper pair, $\lambda_{\cal C}$.
This selection was previously derived for Bogoliubov quasiparticles in multiband BCS superconductors, in terms of the single-particle symmetry under the combined particle-hole-conjugation and inversion operation~\cite{ahn2021theory}.

Let us see how the present approach is related to the previous one in Ref.~\cite{ahn2021theory}.
The Bogoliubov quasiparticle annihilation operator in momentum space has the form
\begin{align}
\hat{\gamma}_{\alpha\bf k}
=(u_{\bf k})_{\alpha\beta}\hat{c}_{\beta\bf k}-(v_{\bf k})_{\alpha\beta}\hat{c}^{\dagger}_{\beta{\bf -\bf k}},
\end{align}
where $\hat{c}_{\beta{\bf \bf k}}$ is the electron annihilation operator, and we use the Einstein summation convention.
$\hat{\gamma}_{\alpha\bf k}$ does not have a well-defined particle number.
Only the number parity is well defined.
Note that we change the notation slightly such that $\alpha$ is the index for both orbital and spin degrees of freedom but does not include momentum, while $\alpha$ includes all labels including momentum or position in Eqs.~\eqref{eq:two-particle} and~\eqref{eq:MF-two-particle}.

In BCS theory, the Bogoliubov quasiparticle creation operator can be written as the single-particle particle-hole conjugation ($P$) of the annihilation operator as
\begin{align}
\hat{\gamma}^{\dagger}_{\alpha-\bf k}
&=-(v^*_{-\bf k})_{\alpha\beta}\hat{c}_{\beta{\bf \bf k}}+(u^*_{-\bf k})_{\alpha\beta}\hat{c}^{\dagger}_{\beta-\bf k}
= (\hat{\gamma}_{\alpha\bf k})^P,
\end{align}
where the single-particle particle-hole conjugation is defined by
\begin{align}
\label{eq:single-particle-PH}
(\hat{\gamma}_{\alpha\bf k})^P
&\equiv
\begin{pmatrix}
c_{\bf k}& c^{\dagger}_{-\bf k}
\end{pmatrix}
U_PK
\begin{pmatrix}
u^T_{-\bf k}\\
-v^T_{-\bf k}
\end{pmatrix}\notag\\
&=
\begin{pmatrix}
c_{\bf k}& c^{\dagger}_{-\bf k}
\end{pmatrix}
\begin{pmatrix}
0&1\\
1&0
\end{pmatrix}
K
\begin{pmatrix}
u^T_{-\bf k}\\
-v^T_{-\bf k}
\end{pmatrix},
\end{align}
where $K$ is the complex conjugation operator.
Therefore, we can write the inversion pair of $\hat{\gamma}^{\dagger}_{\alpha\bf k}$ as
\begin{align}
\hat{I}_{\phi}\hat{\gamma}^{\dagger}_{\alpha\bf k}\hat{I}_{\phi}^{-1}
&=\hat{I}_{\phi}(\hat{\gamma}_{\alpha-\bf k})^P\hat{I}_{\phi}^{-1}
=(\hat{\gamma}_{\alpha\bf k})^{I_{\phi}P},
\end{align}
where $(\hat{\gamma}_{\alpha\bf k})^{I_{\phi}P}=((\hat{\gamma}_{\alpha\bf k})^{P})^{I_{\phi}}$, and
\begin{align}
\label{eq:single-particle-inversion}
(\hat{\gamma}_{\alpha\bf k})^{I_{\phi}}
&\equiv
\begin{pmatrix}
c_{\bf k}& c^{\dagger}_{-\bf k}
\end{pmatrix}
U_{I_{\phi}}
\begin{pmatrix}
u^T_{\bf k}\\
-v^T_{\bf k}
\end{pmatrix},\notag\\
U_{I_{\phi}}
&=
\begin{pmatrix}
e^{i\phi}U^*_I&0\\
0&e^{-i\phi}U_I
\end{pmatrix}
\end{align}
is the single-particle representation of $\hat{I}_{\phi}$ in the Nambu basis $(c_{\bf k}\;c^{\dagger}_{-\bf k})$.
The two-particle state $\ket{\psi_{\alpha}}$ can thus be viewed equivalently as a state with a quasiparticle $\hat{\gamma}^{\dagger}_{\alpha \bf k}$ and an anti-quasiparticle of $(\hat{\gamma}_{\alpha\bf k})^{I_{\phi}P}$.
\begin{align}
\ket{\psi_{\alpha}}=\hat{\gamma}^{\dagger}_{\alpha \bf k}(\hat{\gamma}_{\alpha\bf k})^{I_{\phi}P}\ket{\Omega_{\rm MF}}.
\end{align}
In this picture, the matrix element $\braket{\psi_{\alpha}|\hat{\bf J}|{\Omega}_{\rm MF}}$ describes the transition to the single-particle state $\ket{\alpha \bf k}$ from the single-particle state $U_{I_{\phi}}U_PK\ket{\alpha \bf k}$.

The particle-number conservation can be recovered from the BCS mean-field solution as follows~\cite{tinkham2004introduction}.
We insert the Cooper pair operator in the definition of the Bogoliubov quasiparticle~\cite{josephson1962possible} so that its particle number is well defined.
$
\hat{\gamma}_{\alpha\bf k}
=(u_{\bf k})_{\alpha\beta}\hat{c}_{\beta\bf k}-(v_{\bf k})_{\alpha\beta}\hat{c}^{\dagger}_{\beta{\bf -\bf k}}\hat{\cal C},
$
where
$
\hat{\cal C}^{\dagger}
=\sum_{\bf k}\hat{c}_{\alpha {\bf k}}^{\dagger}(u_{\bf k}^{-1}v_{\bf k})_{\alpha\beta}\hat{c}^{\dagger}_{\beta{-\bf k}}
$
is the Cooper pair creation operator.
The ground state is projected to the $N$-particle state.
$
\ket{{\Omega}^G}
\propto \hat{P}_{N}e^{\hat{\cal C}^{\dagger}}\ket{0}\propto (\hat{\cal C}^{\dagger})^{N/2}\ket{0}.
$
$\ket{{\Omega}^G}$ has a well-defined particle number $N$, and it is an inversion eigenstate when $\hat{\cal C}^{\dagger}$ has a definite inversion parity.

\section{Examples}
\label{sec:examples}

\subsection{Spin-triplet pairing of electrons}

Let us first consider the spin-triplet pairing of electrons (or pairing of spin-polarized fermions) in centrosymmetric systems.
We first consider the spin selection rule.
Spin rotation symmetry around the axis of the Cooper pair spin requires that the two Bogoliubov quasiparticles in $\ket{\psi_{\alpha}}$ carry the same spin whose direction is parallel to the Cooper pair spin. Let us suppose that $\hat{z}$ is the Cooper pair spin direction.
The spin-$z$ selection rule imposes $2s_z^{\gamma}+s_z^{J}-s_z^{\cal C}=2s_z^{\gamma}-s_z^{\cal C}=0$, where $s^O_z$ is the spin of $\hat{O}$ along the $\hat{z}$ direction, because the current operator does not carry spin.
We have $s_z^{\gamma}=1/2$ for triplet pairing.
In this case, a nontrivial inversion selection rule is due to the spatial inversion without spin rotation.
Since $\eta_{\gamma}=-1$ and $\epsilon_I=1$, $\braket{\psi_{\alpha}|\hat{\bf J}|{\Omega}}=0$ for odd-parity pairing $\lambda_{\cal C}=-1$.
Note that while spin-triplet pairing is usually odd-parity pairing, even-parity spin-triplet pairing is not excluded in multiband systems, because the Cooper pair can form an orbital singlet. 

\subsection{Spin-singlet pairing of electrons}

When the Cooper pair is a spin singlet of electrons, an optically created Bogoliubov quasiparticle pair should carry the opposite spin by the spin conservation.
The inversion selection rule is imposed when $\hat{I}$ flip the spin such that $\hat{I}^2=\epsilon_I=-1$: $\hat{I}$ is a combination of spatial inversion and a spin rotation by $\pi$.
Then, $\braket{\psi_{\alpha}|\hat{\bf J}|{\Omega}}=0$ for even-parity pairing $\lambda_{\cal C}=1$.

\subsection{Pairing of spin-orbit coupled electrons}

If the spin-orbit coupling is significant, it is not useful to think about spin rotation symmetry.
We thus consider spatial inversion without spin rotation, satisfying $\epsilon_I=1$, in spin-orbit coupled electronic systems.
Since $\eta_{\gamma}=-1$ and $\epsilon_I=1$, $\braket{\psi_{\alpha}|\hat{\bf J}|{\Omega}}=0$ for odd-parity pairing $\lambda_{\cal C}=-1$.

When time reversal symmetry is present, it is possible to optically excite a pair of quasiparticles with zero total momentum even for odd-parity pairing.
This is because the optical excitation of time-reversal-related pairs is not forbidden:
time-reversed partner of a Bogoliubov quasiparticle is different from the inversion-related partner because time reversal flips spin while inversion does not.
This is in contrast to the systems with spin rotation symmetry, where the time-reversal-related pair is identical to the pair related by an effective inversion -- defined by the combination of spatial inversion and a spin $\pi$ rotation.

\subsection{Charged bosonic excitations with integer spin}

Let us consider the excitation of two charged bosons related by inversion symmetry.
We suppose that the boson carries an integer spin.
Since $\eta_{\gamma}=1$ and $\epsilon_I=1$,
$\braket{\psi_{\alpha}|\hat{\bf J}|{\Omega}}=0$ when $\lambda_{\cal C}=1$.
This can relevant to the optical excitation of Cooperons (pair density wave excitations) in charge-$4e$ superconductors~\cite{kivelson1990doped,berg2009charge}.

\subsection{Superconductivity in doped Mott insulators}

The Mott insulating phase of the Hubbard model has two different quasiparticle-like excitations: singly occupied electronic state called holons and doubly occupied electronic states called doublons, defining the upper and lower Hubbard bands.
Holon and doublon creation operators, $\hat{\zeta}^{\dagger}_{i\sigma}=\hat{c}^{\dagger}_{i\sigma}\hat{n}_{i\bar{\sigma}}$ and $\hat{\eta}^{\dagger}_{i\sigma}=\hat{c}_{i\alpha}^{\dagger}(1-\hat{n}_{i\bar{\sigma}})$, respectively, do not satisfy the full fermionic canonical commutation relations.
However, they satisfy $\{\hat{\zeta}^{\dagger}_{i\sigma_1},\hat{\zeta}^{\dagger}_{j\sigma_2}\}=\{\hat{\eta}^{\dagger}_{i\sigma_1},\hat{\eta}^{\dagger}_{j\sigma_2}\}=0$.
Also, they transform like electrons under symmetry operations.
If these two properties are inherited to the corresponding excitations in the superconducting state, such that Eqs.~\eqref{eq:condition1} and~\eqref{eq:condition2} are satisfied, our selection rule applies to those excitations.
We show in Appendix.~\ref{sec:PYH-model} that a model of superconductivity~\cite{phillips2020exact} in a momentum-space analog of the Hubbard model indeed satisfies these conditions.

\section{Bogoliubov Fermi surfaces}
\label{sec:BFS}

The optical selection rule in superconductors and the stability of the Bogoliubov Fermi surfaces are closely related as we explain in Sec.~\ref{sec:selection-charge-conserving}:
Inversion pairs of Bogoliubov quasiparticles can be excited in the class of superconductors where the Bogoliubov Fermi surfaces are topologically stable~\cite{ahn2021theory}.
Therefore, we can expect unique low-energy optical responses from Bogoliubov Fermi surfaces.
In some symmetry classes, Bogoliubov Fermi surfaces can carry two kinds of topological charges~\cite{agterberg2017bogoliubov,bzdusek2017robust}.
We show in Sec.~\ref{sec:optical-BFS} that doubly charged Bogoliubov Fermi surfaces feature divergent optical responses in the clean-limit, while singly charged Bogoliubov Fermi surfaces feature trivial optical responses.
We first take the conventional BdG formalism to define the topological charges and calculate the optical conductivity, where particle-hole symmetry is crucial for topological charges and the selection rule.
However, as we noted in the introduction and showed in previous sections that we do not need a particle-hole symmetry in the many-body approach.
This implies that topological charges of Bogoliubov Fermi surfaces can also be formulated without particle-hole symmetry.
We demonstrate this in Sec.~\ref{sec:optical-BFS}.

\subsection{Optical conductivity}
\label{sec:optical-BFS}

Bogoliubov Fermi surfaces can carry two topological charges in non-interacting superconductors of ${\bf k}$-local symmetry class D or BDI~\cite{bzdusek2017robust}.
Class D is characterized by the symmetry under the combination of inversion $I$ and particle-hole conjugation $P$, satisfying $(U_{IP}K)^2=1$.
Here, $U_{IP}=U_{I_{\phi}}U_P$, and $U_{I_{\phi}}$ and $U_P$ are matrix representations of the symmetry operators for single-particle excitations [See Eq.~\eqref{eq:single-particle-PH} and~\eqref{eq:single-particle-inversion}.].
There are two topological charges: the Pfaffian invariant (0D charge) and the Chern number (2D charge).
A Bogoliubov Fermi surface is called doubly charged when it carries both kinds of charges.
Class BDI has additional symmetry under $IT,$ satisfying $(U_{IT}K)^2=1$, where $U_{IT}=U_{I_{\phi}}U_T$, and $U_T$ is the single-particle representation of time reversal.
Two topological charges in class BDI are the Pfaffian invariant and the $O(N)$ winding number (1D charge), where $N$ is the number of bands in the normal state.

Below we consider class BDI in two dimensions and class D and BDI in three dimensions.
We note that we use ${\bf k}$-local symmetry classes~\cite{bzdusek2017robust}, which are distinct from the conventional Altland-Zirnbaurer classes defined by symmetries local in real space~\cite{altland1997nonstandard,schnyder2008classification}.

We focus on doubly charged Bogoliubov Fermi surfaces because singly charged ones show negligible optical responses.
Let us note that the two-band single-particle effective Hamiltonian of a singly charged Bogoliubov Fermi surface has the form $H=f({\bf k})\sigma_z$.
In this effective model, no optical excitations occur because the single-particle current operator $J^a=\d H/\d k^a\propto \sigma_z$ has zero off-diagonal component.
Therefore, only additional bands beyond the minimal model can contribute to nontrivial optical responses for singly charged Bogoliubov Fermi surfaces.
In contrast, doubly charged Bogoliubov Fermi surfaces show characteristic nontrivial low-energy optical responses as we show below.

\subsubsection{Two dimensions}

\begin{figure}[t!]
\includegraphics[width=8.5cm]{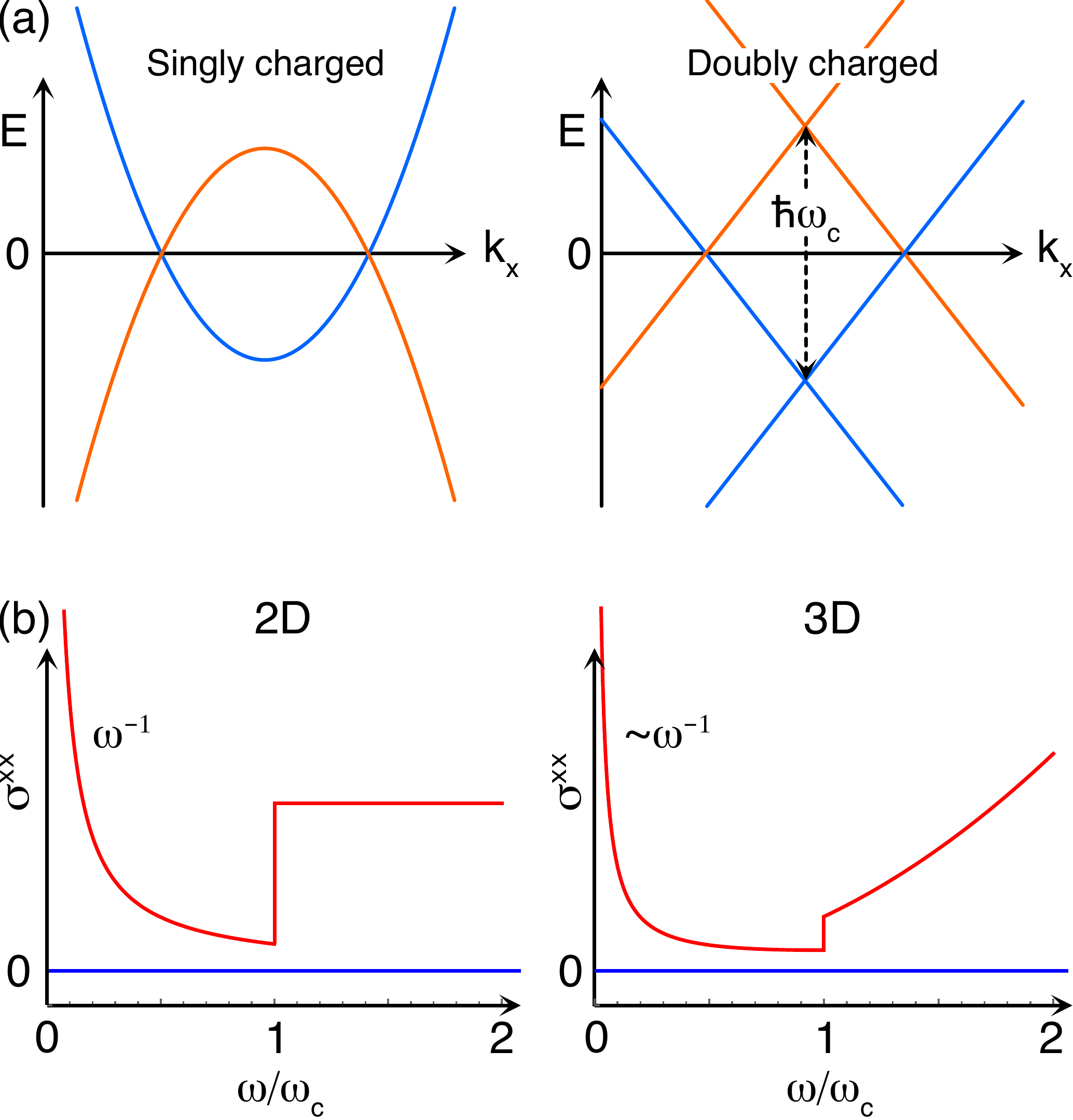}
\caption{
Optical conductivity of singly and doubly charged Bogoliubov Fermi surfaces.
(a) Schematic low-energy single-particle band structure.
We consider isotropic Bogoliubov Fermi surfaces centered at ${\bf k}=0$ and show the band structure along $k_x$ passing through ${\bf k}=0$.
(b) Optical conductivity.
Blue and red curves corresponds to singly and doubly charged cases.
We consider the form $H=f({\bf k})\sigma_z$ for singly charged Bogoliubov Fermi surfaces: $\sigma^{ab}(\omega)=0$ in this case.
We use Eq.~\eqref{eq:BFS-model-2D} and~\eqref{eq:BFS-model-3D} for doubly charged Bogoliubov Fermi surfaces in 2D and 3D, respectively, with $\Delta=0.2\hbar\omega_c$.
$\sigma^{ab}=\sigma^{xx}$ for $b=a$ and $\sigma^{ab}=0$ for $b\ne a$.
}
\label{fig:BFS}
\end{figure}

We consider the simplest model of a doubly charged Bogoliubov Fermi surface in two dimensions, where the single-particle Bogoliubov-de Gennes Hamiltonian has the form
\begin{align}
\label{eq:BFS-model-2D}
H
=\hbar v(k_x\sigma_x+k_y\sigma_z)-\mu\tau_z+\Delta\tau_x.
\end{align}
This Hamiltonian has spacetime inversion $IT$ symmetry $U_{IT}H^*({\bf k})(U_{IT})^{-1}=H({\bf k})$ and spatial-inversion-particle-hole-conjugation $IP$ symmetry $U_{IP}H^*({\bf k})(U_{IP})^{-1}=-H({\bf k})$, where $U_{IT}=1$ and $U_{IP}=\tau_y\sigma_y$.
As the $\sigma$ terms and $\tau$ terms commute with each other, the energy eigenstates are in the form $\ket{n^{\tau,\sigma}}=\ket{n^{\tau}}\otimes\ket{n^{\sigma}}$, where $\ket{n^{\tau}}$ and $\ket{n^{\sigma}}$ are two-component states satisfying
$(-\mu\tau_z+\Delta\tau_x)\ket{n^{\tau}}
=s_{\tau}\sqrt{\mu^2+\Delta^2}\ket{n^{\tau}}$
and $(k_x\sigma_x+k_y\sigma_z)\ket{n^{\sigma}}
=s_{\sigma}\sqrt{k_x^2+k_y^2}\ket{n^{\sigma}}$,
where $s_{\tau}=\pm 1$ and $s_{\sigma}=\pm 1$, such that the energy eigenvalues are
\begin{align}
E^{\tau,\sigma}=s_{\tau} \sqrt{\mu^2+\Delta^2}+s_{\sigma}\hbar v\sqrt{k_x^2+k_y^2}.
\end{align}
Band crossing occurs at zero energy when $k\equiv \sqrt{k_x^2+k_y^2}=\sqrt{\mu^2+\Delta^2}$, which forms a closed loop [Fig.~\ref{fig:BFS}(a)].
We call this the Bogoliubov Fermi surface.

The Bogoliubov Fermi surface is topologically stable because of its nontrivial 0D topological charge protected by $IP$ symmetry.
Since $U_{IP}H^*({\bf k})$ is an anti-symmetric real matrix, its Pfaffian is well defined and real valued.
The explicit expression of the Pfaffian is $\Pf\left[U_{IP}H^*({\bf k})\right]=\sqrt{\mu^2+\Delta^2}-k$ or $-(\sqrt{\mu^2+\Delta^2}-k)$.
There are two possible choices of the sign at each ${\bf k}$.
However, we require that the Pfaffian is continuous in ${\bf k}$, and this allows only one global choice of the sign.
After the global sign is fixed, the sign of the Pfaffian defines a well-defined 0D ${\bb Z}_2$ topological invariant at each ${\bf k}$.
This invariant changes when there is a band crossing at $E=0$~\cite{agterberg2017bogoliubov,bzdusek2017robust}, such that this change defines the 0D topological charge of the band crossing appearing at zero energy, which forms the Bogoliubov Fermi surface.

In addition, the Bogoliubov Fermi surface carries an additional 1D topological charge protected by the simultaneous presence of $IP$ and $IT$ symmetries.
This charge is given by the winding number of the pseudospin ${\bf f}=\hbar v(k_x,k_y)$ for $\sigma$~\cite{bzdusek2017robust}, which is one in our model.
See Ref.~\cite{bzdusek2017robust} for a general definition.
This nontrivial winding number adds stability to the Bogoliubov Fermi surface by protecting it from annihilating itself after being shrunken to a point.

Let us now calculate the real part of the optical conductivity tensor in the clean limit, which has the form
$\sigma^{ab}(\omega)
=\delta^{ab}(\pi/2\hbar\omega)
\int_{\bf k}\sum_{n,m}f_{nm}|J^c_{mn}|^2\delta(\omega-\omega_{mn})$.
It can be calculated simply as the current matrix element is separated for the $\tau$ and $\sigma$ parts.
$J^a_{mn}
=\braket{m|\d_{\bf A}H|n}|_{{\bf A}=0}
=\frac{e}{\hbar}\braket{m^{\tau}|\tau_z|n^{\tau}}\braket{m^{\sigma}|\d_{k^a}H|n^{\sigma}}$.
Here, $\tau_z$ appears because electrons and holes couple to the gauge field ${\bf A}$ differently through ${\bf k}\rightarrow {\bf k}+(e/\hbar){\bf A}$ and ${\bf k}\rightarrow {\bf k}-(e/\hbar){\bf A}$, respectively.
We obtain
\begin{align}
\sigma^{ab}(\omega<\omega_c)
&=
\delta^{ab}\frac{1}{\hbar\omega}\frac{e^2}{4\hbar}
\frac{\Delta^2}{\omega_c}
\end{align}
for $\omega_c=2\sqrt{\mu^2+\Delta^2}/\hbar$, while $\sigma^{ab}(\omega)=\delta^{ab}e^2(16\hbar)^{-1}$ is constant for $\omega>\omega_c$.
The optical conductivity is plotted in the left panel of Fig.~\ref{fig:BFS}(b).

\subsubsection{Three dimensions}

In three dimensions, we begin with a class D Hamiltonian
\begin{align}
\label{eq:BFS-model-3D}
H
=\hbar v(k_x\sigma_x+k_y\sigma_y+k_z\sigma_z)
-\mu\tau_z+\Delta_1\tau_x+\Delta_2\tau_y,
\end{align}
which has $IP$ symmetry under $U_{IP}=\tau_y\sigma_y$.
This Hamiltonian describes a Bogoliubov Fermi surface carrying both the nontrivial Pfaffian invariant and Chern number 2 (note that we have two copies of identical Weyl fermions after we continuously deform the system to have $\mu=\Delta_1=\Delta_2=0$.).
The optical conductivity tensor for $\omega<\omega_c=2\sqrt{\mu^2+\Delta^2}/\hbar$ is
\begin{align}
\sigma^{ab}(\omega<\omega_c)
&=
\delta^{ab}\frac{1}{\hbar\omega}
\frac{e^2}{6\pi\hbar}
\frac{\Delta^2}{\hbar v}
\left[1+\left(\frac{\omega}{\omega_c}\right)^2\right],
\end{align}
where $\Delta=\sqrt{\Delta_1^2+\Delta_2^2}$.
When $\omega>\omega_c$, an additional term $\delta^{ab}e^2(6\pi\hbar)^{-1}(\mu/\omega_c)^2(\omega/v)$ appears.
See the right panel of Fig.~\ref{fig:BFS}(b).

If we impose $IT$ symmetry with $U_{IT}=1$ additionally, then the Hamiltonian reduces to the form we presented for two dimensions.
It describes a cylindrical Bogoliubov Fermi surface.
The optical conductivity of this has the same form as the two-dimensional one and features the low-energy $\omega^{-1}$ divergence again.

\subsection{Alternative definitions of the topological charges}
\label{sec:alternative}

Here we present an alternative expression of the topological invariants of the non-interacting Bogoliubov Fermi surfaces within the mean-field description.
As we are interested in inversion-symmetric systems, we focus on the effective Brillouin zone defined by $k_x\ge 0$.
In the mean-field description, the degrees of freedom at ${\bf k}$ are decoupled from the degrees of freedom at the other momenta and are described by the ${\bf k}$-local Hamiltonian
$
\hat{H}^{\rm MF}({\bf k})
=\hat{c}^{\dagger}_{\alpha\bf k}h_{\alpha\beta\bf k}\hat{c}_{\beta\bf k}
+\frac{1}{2}\hat{c}^{\dagger}_{\alpha\bf k}\Delta_{\alpha\beta}({\bf k})\hat{c}^{\dagger}_{\beta-\bf k}
+\frac{1}{2}\hat{c}_{\alpha-\bf k}\Delta^{\dagger}_{\alpha\beta}({\bf k})\hat{c}_{\beta\bf k}
+({\bf k}\rightarrow -{\bf k}),
$
where the full mean-field Hamiltonian is 
$
\hat{H}^{\rm MF}
=\sum_{{\bf k}:k_x>0}\hat{H}^{\rm MF}({\bf k})+\text{const}.
$

We define topological invariants using the ground state $\ket{\Omega^G_{\bf k}}$ of $\hat{H}^{\rm MF}({\bf k})$.
The zero-, one-, and two-dimensional topological invariants are defined by
\begin{align}
\label{eq:alternative}
\lambda({\bf k})
&=\braket{\Omega^G_{\bf k}|\hat{I}_{\phi}|\Omega^G_{\bf k}},\notag\\
w_1(C)
&=\frac{1}{\pi}\oint_C d{\bf k}\cdot\braket{\Omega^G_{\bf k}|\nabla_{\bf k}|\Omega^G_{\bf k}},\notag\\
C_1(S)
&=\frac{1}{2\pi}\oint_S d{\bf S}\cdot\nabla_{\bf k}\times\braket{\Omega^G_{\bf k}|\nabla_{\bf k}|\Omega^G_{\bf k}},
\end{align}
where $C$ and $S$ indicate a closed loop and a closed surface, respectively.
$\lambda=\pm 1$ is simply the inversion eigenvalue, $w_1=0$ or $1$ (mod $2$) is the normalized Berry phase (equivalently, the first Stiefel-Whitney number~\cite{ahn2018band}), and $C_1$ is the first Chern number.
Here, different from other quantities, $w_1$ for a generic loop $C$ is quantized only when there is additional time reversal symmetry, in which case $C_1=0$.

The topological charges of a Bogoliubov Fermi surface are defined by the change of topological invariants across the Bogoliubov Fermi surface.
As we noted in the main text, the Bogoliubov Fermi surface appears when $\lambda({\bf k})$ changes, because the lowest-energy excited state that preserves inversion symmetry is the two-particle state $\ket{\psi_{\alpha\bf k}}=\gamma^{\dagger}_{\alpha k}(\hat{I}_{\phi}\gamma^{\dagger}_{\alpha k}\hat{I}_{\phi}^{-1})\ket{\Omega^G_{\bf k}}$.
Since $\ket{\psi_{\alpha\bf k}}$ has the inversion parity opposite to $\ket{\Omega^G_{\bf k}}$, only one parameter needs to be adjusted to close the gap between them, and the crossing in the momentum space forms the Bogoliubov Fermi surface.

The change of the Berry phase and Chern number in Eq.~\eqref{eq:alternative} across the Bogoliubov Fermi surface defines the 1D and 2D topological charges.
To see this, note that Bogoliubov Fermi surfaces can be continuously deformed to the Fermi surfaces of the normal state by turning off the superconducting pairing.
Since the Berry phase and the Chern number defines a well-defined topological charge of the Fermi surface (e.g., the Fermi surface of a Weyl point is characterized by the Chern number jump of the ground state), it is enough to show that those topological charges remain well defined (i.e., quantized) when superconductivity comes in.
The quantization of the Chern number of a single state living on a (${\bf k}$-space) closed manifold is well known.
As we show below, the first Stiefel-Whitney number is also quantized by time reversal symmetry.
Therefore, they can serve as a topological charge of the Bogoliubov Fermi surface.

Let us show that $w_1$ is ${\bb Z}_2$-quantized by time reversal symmetry.
Time reversal symmetry of the ground state is represented by $\hat{T}\ket{\Omega^G_{\bf k}}=e^{i\theta_T({\bf k})}\ket{\Omega^G_{\bf k}}$, for some $\theta_T({\bf k})$.
This symmetry gives a constraint on the Berry connection that $\braket{\Omega^G_{\bf k}|\nabla_{\bf k}|\Omega^G_{\bf k}}=\frac{1}{2}\nabla_{\bf k}\theta_T({\bf k})$, and so $w_1(C)=\frac{1}{2\pi}\oint_C\nabla_{\bf k}\theta_{\bf k}$ is an integer-valued winding number.
However, the winding number is defined modulo two because a phase rotation $\ket{\Omega^G_{\bf k}}\rightarrow e^{i\phi_{\bf k}}\ket{\Omega^G_{\bf k}}$ leads to the change $\theta_T({\bf k})\rightarrow \theta_T({\bf k})+2\phi({\bf k})$.
This proves that $w_1(C)=0$ or $1$ (mod $2$).

\section{Discussion}
\label{sec:discussion}

One of the implications of our selection rule is that, in the strong pairing regime, superconductivity is expected to be more robust in centrosymmetric systems than in non-centrosymmetric systems with similar interaction strength and Fermi surfaces.
This is because nontrivial pair-breaking optical conductivity in the non-centrosymmetric systems reduces the superfluid stiffness ${\cal D}^{ab}$, according to the FGT sum rule ${\cal D}^{ab}=\frac{2}{\pi}\int^{\infty}_{0^+} d\omega\left[\sigma^{ab}_{1n}(\omega)-\sigma^{ab}_{1s}(\omega)\right]$, where $\sigma^{ab}_{1n}$ and $\sigma^{ab}_{1s}$ are the real part of the conductivity tensor in the normal and superconducting states, respectively.
Berezinskii–Kosterlitz–Thouless (BKT) transition temperature $T_{\rm BKT}$ is also expected to be reduced because it is proportional to the superfluid stiffness~\cite{nelson1977universal,hazra2019bounds}.

Let us discuss how relevant this is in the superconductivity of twisted layered materials~\cite{cao2018unconventional,liu2019spin,park2021tunable,hao2021electric,chen2019signatures,tsai2019correlated,wang2020correlated,an2020interaction,xu2021tunable,balents2020superconductivity}.
Robust superconductivity has been observed in twisted bilayer and trilayer graphene, having twofold rotation symmetry, but not in other twisted materials lacking twofold rotation and inversion symmetries.
When we assume the ratio $\Delta_{\rm BCS}/E_F\sim 0.3$ of twisted trilayer graphene~\cite{park2021tunable,hao2021electric}, where $\Delta_{\rm BCS}= 1.76k_BT_c$, and $T_c\sim T_{\rm BKT}$, the relative enhancement of the superfluid stiffness by the inversion selection rule is a factor of $(\Delta_{\rm BCS}/E_F)^2\sim 0.1$, which is non-negligible but still small.
Nonetheless, it is worth noting that $(k_F\xi)^2\sim 1\gg (\Delta_{\rm BCS}/E_F)^2$ in the optimal regime, where $\xi$ is the superconducting (Ginzburg-Landau) coherence length.
This implies that the superconducting gap is not given by $2\Delta_{\rm BCS}$ but by another pseudogap of scale $\sim\hbar v_F\xi^{-1}$, which is a characteristic of the BCS-BEC crossover regime~\cite{nakagawa2021gate,chen2005bcs,randeria2014crossover}.
If the creation of inversion-related quasiparticle pairs depends on the pseudogap rather than $\Delta_{\rm BCS}$, the inversion selection rule can significantly enhance $T_{\rm BKT}$ in centrosymmetric twisted materials compared to noncentrosymmetric ones.

The same argument applies to iron-based superconductors~\cite{lubashevsky2012shallow,hashimoto2020bose,shibauchi2020exotic} and Li$_x$ZrNCl~\cite{nakagawa2021gate}, which are other few exceptional ultra-strong-coupling superconductors.
They are also in the BCS-BEC crossover regime with a large pseudogap and $k_F\xi\sim 1$.
Hence, the effect of the selection rule are expected to be significant.
Since these materials have inversion symmetry, the suppression of the superconducting optical conductivity by the selection rule contributes to the robustness of the superconductivity

It is worth noting that, in the strong-coupling regime, another contribution of the same order $\propto (\Delta/E_F)^2$ becomes relevant to the superfluid stiffness: the quantum geometric contribution by multiband effects~\cite{peotta2015superfluidity,liang2017band,verma2021optical,chen2021probing,ahn2021superconductivity}.
While the pair breaking reduces the superfluid stiffness, the quantum geometry enforces the superfluid stiffness.
Interplay between the two competing effects will be particularly important in the stability of non-centrosymmetric strong-coupling superconductors.

A large spectral weight in the superconducting state can cause an instability even in centrosymmetric systems, when the inversion-related pair creation of Bogoliubov quasiparticles is allowed by multi-orbital effects (i.e., when $\lambda_{\cal C}=\epsilon_I$).
In this class of systems, Bogoliubov Fermi surfaces (namely, the Fermi surface of Bogoliubov quasiparticles) are topologically protected~\cite{agterberg2017bogoliubov,bzdusek2017robust}.
Because of this topological protection, there remains small Fermi surface pockets generically when the superconducting gap opens.
As the pairing strength becomes large enough, the increase of the optical spectral weight in the superconducting state at low energies cause the negativity of the superfluid stiffness~\cite{wan2009pairing,setty2020topological,setty2020bogoliubov}, making it unstable towards an inversion-broken state.

Furthermore, our mean-field calculations show that, when the Bogoliubov Fermi surface carries two topological charges, the superfluid stiffness is always negative in the clean limit because the spectral weight diverges logarithmically as the temperature approaches zero in the superconducting state (because $\sigma(\omega)\propto \omega^{-1}$ at low frequencies and the spectrum is gapless.).
This is consistent with other studies of inversion-breaking instabilities of Bogoliubov Fermi surfaces~\cite{oh2020instability,tamura2020electronic,timm2021distortional} from different perspectives.
In particular, Ref.~\cite{timm2021distortional} showed that there {\it always} is an inversion-breaking instability, analogous to the Cooper instability for superconductivity.
While it was not emphasized, the models used in those works host doubly charged Bogoliubov Fermi surfaces, which carry both the Pfaffian invariant and the Chern number.

This instability suggests a new type of topological superconductors, having no topologically protected gapless surface states.
It was recently shown in Ref.~\cite{ahn2021unconventional} that doubly charged Bogoliubov Fermi surfaces can appear as anomalous surface states on the twofold-rotation-invariant surfaces of three-dimensional topological crystalline superconductors.
They appear when the superconducting order parameter preserves time reversal symmetry and changes its sign under the twofold rotation.
Our discussion above implies that the surface state is unstable towards the breaking of twofold rotational symmetry --- an effective two-dimensional inversion symmetry on the surface.
As the spontaneous symmetry breaking occurs only on the surfaces, the nontrivial bulk topology and the gapless helical hinge states remain stable.
These properties define novel topological crystalline superconductivity characterized by a spontaneous symmetry breaking on the surface.

\begin{acknowledgments}
J.A. was supported by the Basic Science Research Program through the National Research Foundation of Korea (NRF) funded by the Ministry of Education (Grant No. 2020R1A6A3A03037129) and by the Center for Advancement of Topological Semimetals, an Energy Frontier Research Center funded by the US Department of Energy Office of Science, Office of Basic Energy Sciences, through the Ames Laboratory under contract No. DE-AC02-07CH11358.
N.N. was supported by JST CREST Grant Number JPMJCR1874 and JPMJCR16F1, Japan, and JSPS KAKENHI Grant number 18H03676.
\end{acknowledgments}

\appendix

\section{Philips-Yao-Huang (PYH) model}
\label{sec:PYH-model}

To test our selection rule for unconventional superconductivity in a non-Fermi-liquid metal, we consider the mean-field Philips-Yao-Huang (PYH) model~\cite{phillips2020exact}.
Here we explicitly demonstrate the selection rule by calculating the optical conductivity through the exact diagonalization.
The mean-field Hamiltonian is
\begin{align}
\hat{H}^{\rm MF}
=\sum_{{\bf k}:k_x>0}\hat{H}_{\rm PYH}({\bf k})+\text{const}.
\end{align}
where $\Delta$ is an $s$-wave singlet pairing order parameter, and
\begin{align}
\hat{H}_{\rm PYH}({\bf k})
&=\xi_{\bf k}(\hat{n}_{\uparrow \bf k}+\hat{n}_{\downarrow \bf k}+\hat{n}_{\uparrow -\bf k}+\hat{n}_{\downarrow -\bf k})\notag\\
&+U(\hat{n}_{\uparrow \bf k}\hat{n}_{\downarrow \bf k}+\hat{n}_{\uparrow -\bf k}\hat{n}_{\downarrow -\bf k})\notag\\
&+\Delta(\hat{b}^{\dagger}_{\bf k}+\hat{b}^{\dagger}_{-\bf k})+\Delta^*(\hat{b}_{\bf k}+\hat{b}_{-\bf k}).
\end{align}
is a ${\bf k}$-dependent operator that can be represented as a $16\times 16$ matrix acting on the 16 basis states
\begin{align}
\begin{pmatrix}
1\\
\hat{c}_{\uparrow\bf k}^{\dagger}\\
\hat{c}_{\downarrow\bf k}^{\dagger}\\
\hat{c}_{\uparrow\bf k}^{\dagger}\hat{c}_{\downarrow\bf k}^{\dagger}
\end{pmatrix}
\otimes
\begin{pmatrix}
1\\
\hat{c}_{\uparrow-\bf k}^{\dagger}\\
\hat{c}_{\downarrow-\bf k}^{\dagger}\\
\hat{c}_{\uparrow-\bf k}^{\dagger}\hat{c}_{\downarrow-\bf k}^{\dagger}
\end{pmatrix}
\ket{0}.
\end{align}
The normal state ($\Delta=0$) of this model corresponds to the Hatsugai-Kohmoto model~\cite{hatsugai1992exactly}.

In the superconducting state, the analogs of the Bogoliubov quasiparticle annihilation operators have the form~\cite{phillips2020exact}
\begin{align}
\hat{\gamma}^l_{\sigma\bf k}
&\propto x_{\bf k}\hat{\zeta}_{\sigma\bf k}-\sigma z_{\bf k}\hat{\zeta}^{\dagger}_{\bar{\sigma}-\bf k},\notag\\
\hat{\gamma}^u_{\sigma\bf k}
&\propto z_{\bf k}\hat{\eta}_{\sigma\bf k}-\sigma y_{\bf k}\hat{\eta}^{\dagger}_{\bar{\sigma}-\bf k},
\end{align}
where $\hat{\zeta}^{\dagger}_{\sigma \bf k}=\hat{c}^{\dagger}_{\sigma\bf k}(1-\hat{n}_{\bar{\sigma} \bf k})$ and $\hat{\eta}^{\dagger}_{\sigma \bf k}=\hat{c}^{\dagger}_{\sigma\bf k}\hat{n}_{\bar{\sigma} \bf k}$ are the holon and doublon creation operators for the lower and upper Hubbard bands in the normal state, respectively.
These quasiparticle annihilation operators satisfy the anticommutation relations
\begin{align}
\{\hat{\gamma}^{l}_{\sigma_1{\bf k}_1},\hat{\gamma}^{l}_{\sigma_2{\bf k}_2}\}
&=\{\hat{\gamma}^{u}_{\sigma_1{\bf k}_1},\hat{\gamma}^{u}_{\sigma_2{\bf k}_2}\}
=0,
\end{align}
while other commutation relations (such as the one between $\hat{\gamma}$ and $\hat{\gamma}^{\dagger}$) are not fermionic canonical commutation relations.
Here, $i=l,u$ is the index indicating the lower ($l$) or upper ($u$) Hubbard band, and $\sigma=\uparrow,\downarrow$ is the spin index.
Also, they quasiparticles transforms under inversion as
\begin{align}
\hat{I}\hat{\gamma}^{i\dagger}_{\sigma\bf k}\hat{I}^{-1}
=\hat{\gamma}^{i\dagger}_{\sigma,-\bf k}
\end{align}
for $i=l,u$.
Accordingly, $\eta_{\gamma}=-\epsilon_I=-1$.

The ground state and all the excited states can be simply calculated by diagonalizing the $16\times 16$ matrix.
For concreteness, we consider a one-dimensional system with $\xi_{\bf k}=-\mu-2t\cos k$ and take $\mu=t=\Delta=1$ and $U=16t$.
When $\Delta=0$, the normal state is a hole-doped Mott insulator that has an extensive ground state degeneracy: spin degeneracy at each k.
After we turn on $\Delta\ne 0$, the unique superconducting ground state is chosen by the preferred singlet formation.

We find that both of the ground state and the lowest excited two-particle state of $\hat{H}_{\rm PYH}({\bf k})$ at each ${\bf k}$ have even parity under inversion, so the current matrix element between them should vanish.
This is consistent with our selection rule that an inversion-related pair of quasiparticles is not optically excited when $\eta_{\gamma}=-1$ and $\epsilon_I=1$ for even-parity pairing.
One can directly see the vanishing of the relevant matrix element by using the current operator
$\hat{\bf J}
=\frac{e}{\hbar}
\sum_{\bf k}\frac{\d\xi_{\bf k}}{\d{\bf k}}\left(\hat{n}_{\uparrow \bf k}+\hat{n}_{\downarrow \bf k}\right)$.


%

\end{document}